\documentclass[twocolumn,prb, aps,superscriptaddress]{revtex4}

\usepackage{mathptmx}
\usepackage{subfigure}
\usepackage{dcolumn}
\usepackage{amsmath,amssymb}
\usepackage{bm}
\usepackage{color}
\usepackage{latexsym}
\usepackage{epstopdf}
\usepackage{color}
\usepackage[english]{babel}
\usepackage{latexsym}

\usepackage{psfrag,graphicx} 
\usepackage{epsf} 
\usepackage{subfigure} 
\usepackage{amsmath} 
\usepackage{amssymb} 
\usepackage{amsfonts}
\usepackage{bm}
\usepackage{natbib}
\usepackage{epstopdf}
\usepackage{appendix}
\usepackage[colorlinks=true,citecolor=blue,linkcolor=blue]{hyperref}

\newcommand{\ket}[1]{\ensuremath{|#1\rangle}}
\newcommand{\braket}[1]{\ensuremath{\left\langle{#1}\right\rangle}}



\begin{document}

\title{Statistically induced Phase Transitions and Anyons in 1D Optical Lattices}

\author{Tassilo Keilmann$^*$}
\affiliation{Physics Department and Arnold Sommerfeld Center for Theoretical Physics,
Ludwig-Maximilians-Universit\"at M\"unchen, D-80333 M\"unchen, Germany}
\affiliation{Max-Planck-Institut f\"{u}r Quantenoptik, Hans-Kopfermann-Str. 1, D-85748 Garching, Germany}

\author{Simon Lanzmich}
\affiliation{Physics Department and Arnold Sommerfeld Center for Theoretical Physics,
Ludwig-Maximilians-Universit\"at M\"unchen, D-80333 M\"unchen, Germany}

\author{Ian McCulloch}
\affiliation{Physics Department,
The University of Queensland,
Brisbane, QLD, AUS-4072, Australia}

\author{Marco Roncaglia}
\affiliation{Physics Department and Arnold Sommerfeld Center for Theoretical Physics,
Ludwig-Maximilians-Universit\"at M\"unchen, D-80333 M\"unchen, Germany}
\affiliation{Max-Planck-Institut f\"{u}r Quantenoptik, Hans-Kopfermann-Str. 1, D-85748 Garching, Germany}
\affiliation{Dipartimento di Fisica del Politecnico, corso Duca degli Abruzzi 24, I-10129 Torino, Italy}

\maketitle

\subsection*{Abstract}
\textbf{
Anyons -- particles carrying fractional statistics that interpolate between bosons and fermions -- have been conjectured to exist in low dimensional systems. In the context of the fractional quantum Hall effect (FQHE), quasi-particles made of electrons take the role of anyons whose statistical exchange phase is fixed by the filling factor.
Here we propose an experimental setup to create anyons in one-dimensional lattices with fully tuneable exchange statistics. In our setup, anyons are created by bosons with occupation-dependent hopping amplitudes, which can be realized by assisted Raman tunneling. The statistical angle can thus be controlled in situ by modifying the relative phase of external driving fields. This opens the fascinating possibility of smoothly transmuting bosons via anyons into fermions and of inducing a phase transition by the mere control of the particle statistics as a free parameter.
In particular, we demonstrate how to induce a quantum phase transition from a superfluid into an exotic Mott-like state where the particle distribution exhibits plateaus at fractional densities. 
}

\subsection*{Introduction}

Usually, every particle in quantum theory is classified as either a boson -- a particle joining any number of identical particles in a single quantum state -- or a fermion, characterized by the sole occupancy of its state. 
The exchange of two fermions leads due to the Pauli principle to a phase factor -1 in the total wavefunction, while the wavefunction of two bosons remains invariant under particle exchange.
More than 30 years ago, researchers proposed a third fundamental category of particles living in two-dimensional (2D) systems, ``anyons'' \cite{anyon1,anyon2,anyon3,Tsui, Laughlin}.  For two anyons, the wavefunction acquires a fractional phase $e^{i \theta}$ under particle exchange, giving rise to \emph{fractional statistics}, with $0<\theta<\pi$. 
For a few years the physics of anyons remained restricted to the
2D world\cite{Girvin}, until Haldane presented the concept of fractional
statistics in arbitrary dimensions\cite{Haldane}. 

\begin{figure} []
  \centering
  \includegraphics[width=1\linewidth]{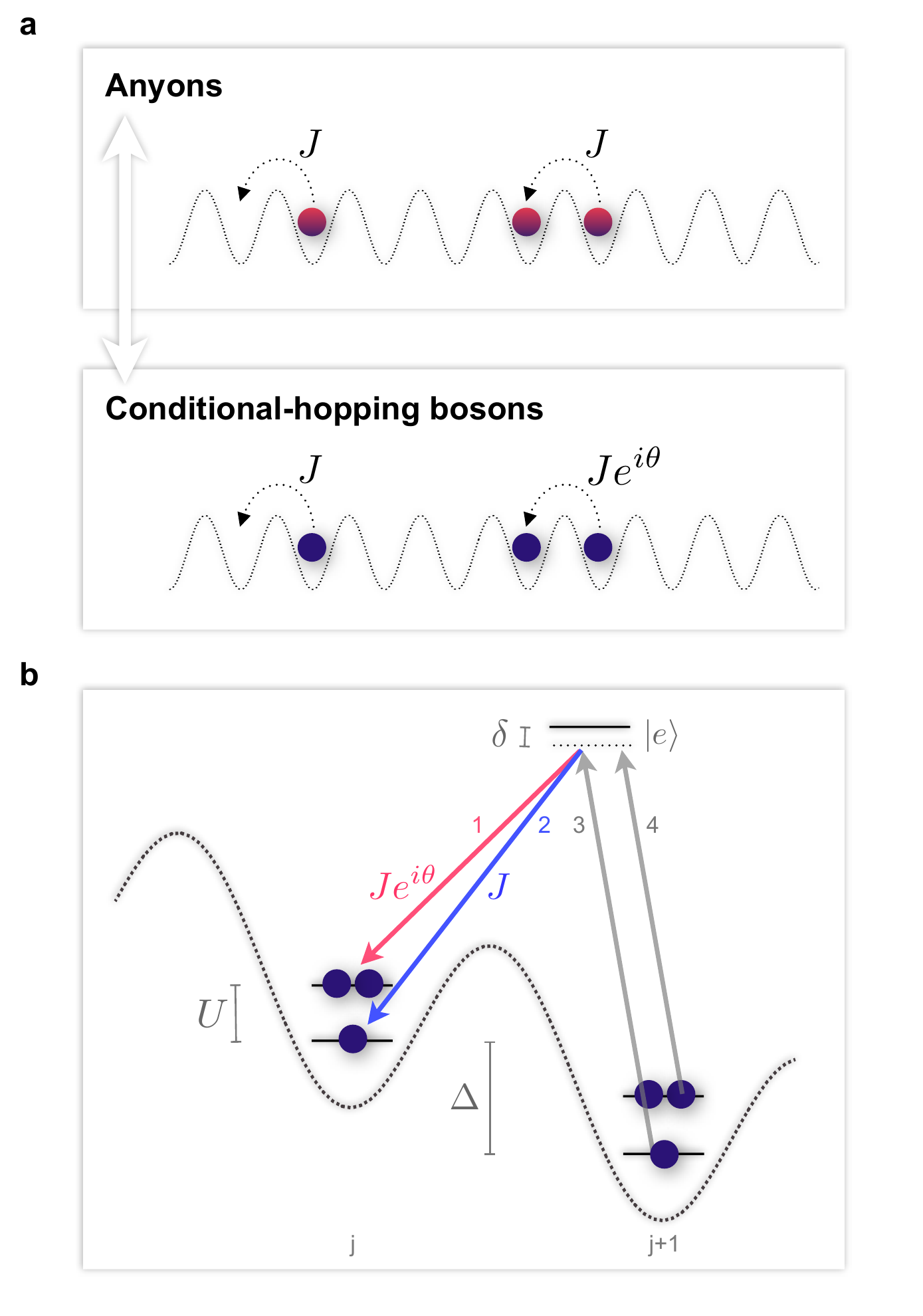}
  \vspace{-0.5cm}
  \caption{
  \textbf{Anyon-Boson Mapping and schematic of the proposed experiment.} (a) Anyons in 1D can be mapped onto bosons featuring occupation-dependent hopping amplitudes.  (b) Assisted Raman tunneling can selectively address hopping processes connecting different occupational states and induce a relative phase, realizing a fully tuneable particle exchange statistics angle $\theta$. Energies are not in scale.
}
  \label{fig:scheme}
\end{figure}

Anyons in one dimension (1D) are still unexplored to a wide extent.
Recently, it has been put forward to create fractional statistics in a 1D
Hubbard model of fermions with correlated hopping processes\cite{bond-charge}. Anyons are realized in this case as low-energy elementary excitations. 

In this Article, we introduce an exact mapping between anyons and bosons in 1D. We show that a Hubbard model of anyons is equivalent to a variant of the Bose-Hubbard model\cite{BHM} where the bosonic hopping amplitudes are state-dependent. This conditional-hopping phase factor breaks reflection parity in the system, which is an important ingredient to realize fractional statistics\cite{Wilczek}. We propose to realize bosons with conditional-hopping amplitudes using assisted Raman tunneling in an optical lattice\cite{Ignacio,Greiner} (OL). We discuss how the direct control of the statistical phase can induce a quantum phase transition from a bosonic superfluid into a Mott-like state, exhibiting exotic Mott shells at fractional densities. The ``statistical ramp'' transmutes bosons smoothly into ``pseudofermions'', with anyons as intermediate steps.

Anyons in 1D are defined\cite{Batchelor, Kundu} by the generalized
commutation relations \begin{equation}
a_{j}a_{k}^{\dagger}-e^{-i\theta\mathrm{sgn}(j-k)}a_{k}^{\dagger}a_{j}=\delta_{jk},\hspace{1em}a_{j}a_{k}=e^{i\theta\mathrm{sgn}(j-k)}a_{k}a_{j},\label{CCR}\end{equation}
where the operators $a_{j}^{\dagger},a_{j}$ create or annihilate
an anyon on site $j$, respectively. The sign function is such that
$\mathrm{sgn}(j-k)=0$ for $j=k$, hence two particles on the
same site behave as ordinary bosons. In consequence, anyons with statistics
$\theta=\pi$ are pseudofermions:  while being bosons on-site, they are fermions off-site.

\subsection*{Results}

\subsubsection*{Mapping between Anyons and Bosons}

We introduce an exact mapping between anyons and bosons in 1D.
Let us define the \emph{fractional version} of a Jordan-Wigner transformation,
\begin{equation}
a_{j}=b_{j}\exp\left(i\theta\sum_{i=1}^{j-1}n_{i}\right)\label{mapping}\end{equation}
with $n_{i}=a_{i}^{\dagger}a_{i}=b_{i}^{\dagger}b_{i}$ the number
operator for both particle types. Provided that the particles of type
$b$ are bosons, $[b_{j},b_{i}^{\dagger}]=\delta_{ji}$ and $[b_{j},b_{i}]=0$,
we can show that the mapped operators $a$ indeed obey the anyonic
commutation relations as introduced in equation (\ref{CCR}). For a proof see Methods. This mapping
elucidates that anyons in 1D are indeed non-local quasi-particles, made of bosons with an attached string operator.

Our ultimate goal is to propose a realistic setup for demonstrating
an interacting gas of anyons in 1D OLs. We therefore introduce the Anyon-Hubbard model 
\begin{equation}
H^{a}=-J\sum_{j}^L (a_{j}^{\dagger}a_{j+1}+\mathrm{h.c.})+\frac{U}{2}\sum_{j}^L n_{j}(n_{j}-1),\label{AHM}\end{equation}
where $J$ is the tunneling amplitude connection two neighbouring sites
and $U$ is the on-site interaction energy. By inserting the Anyon-Boson
mapping, equation (\ref{mapping}), the Hamiltonian $H^{a}$ can be rewritten
in terms of bosonic operators: \begin{equation}
H^{b}=-J\sum_{j}^L (b_{j}^{\dagger}b_{j+1}e^{i\theta n_{j}}+\mathrm{h.c.})+\frac{U}{2}\sum_{j}^L n_{j}(n_{j}-1).\label{bosons}\end{equation}

The mapped, bosonic Hamiltonian thus describes bosons with a occupation-dependent amplitude $Je^{i\theta n_{j}}$ for hopping processes from right to left ($j+1 \rightarrow j$). If the target site $j$ is unoccupied,
the hopping amplitude is simply $J$. If it is occupied by one boson,
the amplitude reads $Je^{i\theta}$, and so on. The conditional hopping scheme is depicted in Fig. 1a.
We emphasize that the non-local mapping between anyons and bosons, equation (\ref{mapping}), leads luckily to a purely local, and thus viable Hamiltonian.

As expected from anyons, the reflection parity symmetry is broken\cite{Wilczek} at the level of the commutation relations, equation (\ref{CCR}).
The fractional Jordan-Wigner transformation, equation (\ref{mapping}), transfers this asymmetry also to the bosonic case:
The resulting Hamiltonian, equation (\ref{bosons}), features a phase factor acting only on the left site $j$ and thus violates parity.
Even in the absence of the on-site interaction, $U=0$, the exponential operator in equation (\ref{bosons}) gives rise to many-body interactions, as expected for anyons\cite{Wilczek}.

In the limit $U/J\to\infty$, bosons are impenetrable and each site
contains at most one particle. In this case, the phase factor disappears
and the bosonic Hamiltonian $H^{b}$ reduces to an ordinary Tonks-Girardeau
gas\cite{Tonks1,Tonks2,Tonks3,Tonks4}. However, we consider local occupation
numbers beyond the hard-core limit, $n_{j}>1$. Thus anyons can exchange
positions, changing the phase of the total wavefunction, and show
non-trivial features. 

\subsubsection*{Experimental realization}

In our experimental proposal, the key point for realizing anyonic statistics is
to induce a hopping term with a phase shift which depends on the
occupation of the left-hand site $j$, Fig.~1b displays the basic concept. In order to distinguish between different local occupational states, we require a non-zero on-site interaction $U$.
For simplicity, let us consider lattice site occupations that are restricted to $n_j=0,1,2$ (higher local truncations are also possible, see Methods). The occupation-dependent tunneling and thus the conditional-hopping model equation (\ref{bosons}) can be implemented in OLs with present experimental techniques.
We propose to employ an assisted tunneling scheme, based on ideas by Jaksch and Zoller\cite{jaksch03} and Juzeliunas \emph{et al.}\cite{fleischhauer}.
The OL is tilted, with an energy offset $\Delta$ between neighbouring sites, this additional field gradient breaks reflection parity. 
Two different occupational states (note that the occupational state $n_j=0$ is not relevant) in either of the two sites form in total a 4-dimensional atomic ground state manifold, which we propose to couple to an excited state $|e\rangle$ via four external driving fields (labeled 1, 2, 3 and 4 in Fig. 1b). According to this notation, singly- and doubly-occupied states are
 coupled by fields 2 and 1 in the left site and by 3 and 4 in the right site, respectively. 
The excited state can be experimentally realized in at least two alternative ways.\\
First, two spin-dependent lattices\cite{Mandel, Campbell, McKay} can be employed. In the case of Rb$^{87}$, one lattice for example traps atoms in the $F=1, m_F=-1$ hyperfine state, assigned to the ground state manifold. The excited state $|e\rangle$ can then be engineered as a vibrational state of a second lattice, trapping atoms in the $F=1, m_F=0$ hyperfine state. Atoms in the excited state would then be localized between the left and right wells of the $F=1, m_F=-1$ lattice, but not necessarily in their center.\\
This implementation would offer the advantage of external driving fields in the radio-frequency regime. Such frequencies could then still resolve\cite{Campbell} the typical energies $U$ and $\Delta$ (both of the order of a few kHz), which is necessary to selectively couple to the four different states in the ground state manifold.\\
Second, one can use two optical lattices, and trap ground state manifold atoms in the red-detuned lattice, while the excited state would live in the blue-detuned one. The driving fields necessary in this case would be then, however, typically in the THz frequency range, making a precise resolution of $U$ and $\Delta$ more challenging for the experimentalist: in principle a laser with a linewidth $\delta_{\textrm{linewidth}}  \ll U, \Delta$ would be needed.\\
The effective tunneling rates $J_{ab}$ ($a \in \{1,2\}, b \in \{3,4\}$) are calculated for four $\Lambda$-systems, laser frequencies $\omega_i$, and Rabi frequencies $\Omega_i$ ($i \in \{1,2,3,4\}$) via adiabatic elimination, see Methods. We emphasize that the tilt energy $\Delta$ disappears in the effective Hamiltonian after rotating out time-dependent phase factors: indeed this energy offset is absorbed (or released) by the external radiation field, yielding a total Hamiltonian without a tilt term (see also Jaksch and Zoller\cite{jaksch03} on this issue).  \\
The following conditions on the effective tunneling rates $J_{ab}$ have to be satisfied in order to realize our model in equation (\ref{bosons}):
\begin{eqnarray}
J_{23} &=& J_{24} \equiv J \, ,\label{condition1}  \\ 
J_{13} &=& J_{14} \equiv J e^{i \theta} \label{condition2} \, , 
\end{eqnarray}
where $\theta$ is the anyonic exchange phase.
For a more detailed consideration of realistic energy scales and appropriate parameter regimes, see Methods.

\begin{figure} []
  \centering
  \includegraphics[width=1.05\linewidth]{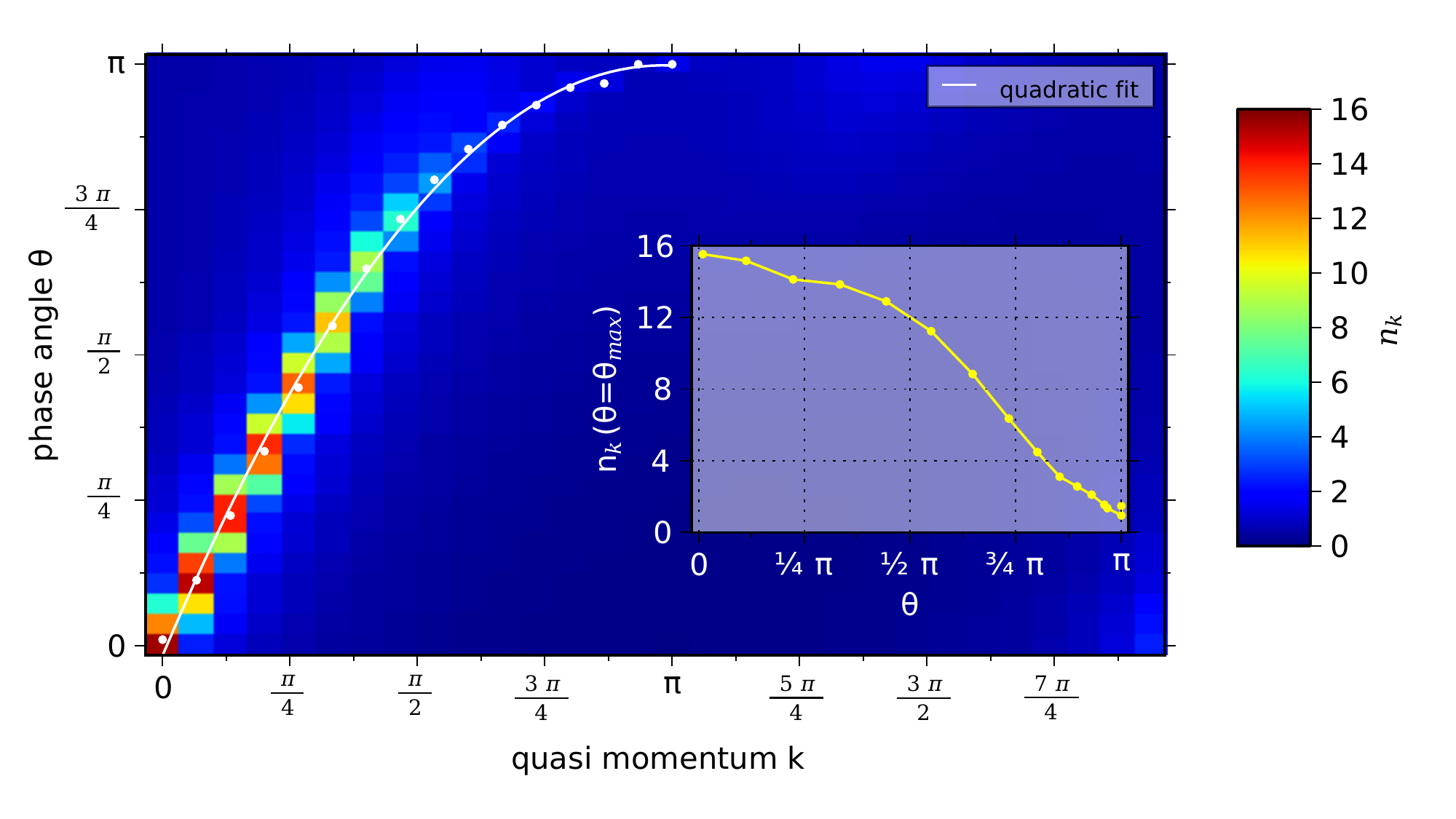}
  \vspace{-0.5cm}
  \caption{
  \textbf{Density distribution in quasi-momentum space \braket{n_k} as a function of particle statistics $\theta$.}  The shift of the density peaks with increasing $\theta$ displays a characteristic quadratic behaviour. The  fit to the trace of density maxima is depicted in white and yields fitting parameters $k_0= 0.9828 \pi$, $\alpha= -1.0511/ \pi$, $\beta= 0.9982 \pi$. The inset displays the decrease of the peak occupations with $\theta$ (yellow circles) and indicates the \emph{statistically induced} phase transition from a superfluid to a Mott-like state. Parameters: $L=30, N=31, U/J=0.2$.
}
  \label{fig:nk}
\end{figure}

\subsubsection*{Density in quasi-momentum space}

We have computed the ground state wave function for the conditional-hopping Bose-Hubbard model, equation (\ref{bosons}), using the Density Matrix Renormalization Group (DMRG)\cite{White,DMRG}.
In Fig. 2 we plot the quasi-momentum distribution
\begin{equation}
\langle n_k \rangle = \frac{1}{L} \sum_{ij} e^{ik(x_i-x_j)} \langle b_{i}^\dagger b_{j}\rangle
\end{equation}
as a function of the statistical phase angle $\theta$. The case $\theta = 0$ corresponds to ordinary bosons, which for the parameters chosen quasi-condense. The density distribution in quasi-momentum space is thus peaked at $k=0$. Increasing $\theta$ to non-zero values, we find that the position of the peaks $\theta_{max}(k)$ is shifted as a non-linear function of $k$. Indeed, for fillings $N/L>1$ one finds a quadratic behaviour $\theta_{max}(k) = \alpha (k-k_0)^2 + \beta$. For fillings close to $N/L=1$, $\beta\rightarrow \pi, k_0 \rightarrow \pi$ and $\alpha \rightarrow -1/\pi$. For higher fillings $N/L\rightarrow 2$, these fitting parameters are altered, however, the characteristic quadratic dependence is conserved.\\
In this analysis, we find two important characteristics of conditional hopping bosons and thus anyons in 1D. The quadratic dependence of $\theta_{max}$ on $k$ contrasts ordinary magnetic fields (with a constant phase factor $e^{i \theta}$ in the kinetic Hamiltonian). In this case, the shift of the peaks depends linearly on the phase angle $\theta$. 
In the anyonic case, however, the growth of correlations with increasing $\theta$ may induce the characteristic quadratic trace, which could be directly observed in OL experiments using standard time-of-flight imaging.

An even more important observation is that the contrast of the peaks (and the phase coherence) decays with increasing $\theta$. The peak values $\braket{n_k(\theta=\theta_{max})}$ are plotted in the inset of Fig. 2, in the pseudofermionic limit ($\theta \rightarrow \pi$) the peak is almost washed out. This suggests that an increase of $\theta$ transforms the initial quasi-condensate into a quantum state where all phase coherence is lost. It will become evident in the subsequent paragraphs that this quantum state will turn out to be a Mott-like state, with Mott plateaus emerging at fractional densities. We emphasize that this quantum phase transition is only driven by the statistical angle $\theta$, all other parameters such as $J/U$ are fixed. The loss of coherence can be understood as follows. With increasing $\theta$ the occupation-dependent phase factor in equation (\ref{bosons}) becomes more and more important: Hopping processes connecting sites with different occupations will contribute different phases and will cancel out in the kinetic Hamiltonian due to incoherent superpositions. This destructive interference effect is amplified by an increasing $\theta$ and induces the localization of particles, yielding an insulating phase.
We emphasize that the present analysis of the density distribution in momentum space refers to the bosonic particles only, $\langle n_k \rangle=\langle b_k^{\dagger} b_k \rangle$. Namely, Fig. 2 represents what would be observed in the experiment. However, while the mapping in equation (\ref{mapping}) establishes a 1-1 correspondence between the number operators in real space, $\langle a_j^{\dagger}a_j\rangle=\langle b_j^{\dagger}b_j\rangle$, the density distributions in momentum space may differ significantly. In this sense, a study of $\tilde{n}_k=\langle a_k^{\dagger}a_k\rangle$ and the superfluid order parameter in the original anyonic model (\ref{AHM}) would be very interesting, but it is beyond the scope of this paper. A study of density distributions in momentum and real space was recently presented for the particular case of hard-core anyons\cite{Hao}.

\subsubsection*{Phase diagram}

Next, we present the phase diagram for conditional-hopping bosons in the $(\mu, J/U)$-plane.  Just as in the case of ordinary bosons ($\theta=0$) -- the celebrated phase diagram of the Bose-Hubbard model\cite{BHM} -- the anyonic version will display insulating and superfluid regions. However, we find that the size of the insulating regions (Mott lobes) grows with increasing statistical angle $\theta$ -- a fact that will be central in our proposal to design phase transitions by tuning the particle statistics. \\
The phase diagram is calculated as follows. We start with a unit filling ground state $\ket{N=L}$ and compare the energies with the ground states of $\ket{N=L\pm1}$. This yields the gap energies $\Delta E^{\pm} (L)$, corresponding to the energy required to add or subtract one particle, respectively. These gap energies were calculated using DMRG for system sizes $L=15, 30, 60, 90$ and $120$. From the finite-size scaling we extrapolate the infinite-size values $\Delta E^{\pm}$, which are plotted in Fig. 3 in the $(\mu, J/U)$-plane. Note that a Mott insulator is associated with a non-zero gap $\epsilon = \Delta E^+ - \Delta E^-$, while for the superfluid phase $\epsilon =0$ in the thermodynamic limit. For the bosonic case $\theta=0$ we recover the first Mott lobe as in the work by K\"uhner and Monien\cite{BHM1D}. Turning on the statistical angle and transmuting the bosons into anyons, we observe an expansion of the Mott-like insulating phase in both dimensions $\mu$ and $J/U$. The phase transition points $(U/J)_{crit}$ (defined by the cusps of the Mott lobes) are plotted in the inset of Fig. 3 as a function of $\theta$.
In the pseudofermion limit  $\theta \rightarrow \pi$, the Mott-like phase seems to extend to very large values of $J/U$. In contrast, ordinary bosons would form a superfluid state in this parameter regime.
The expansion of the Mott lobes with $\theta$ is also observed in mean-field calculations for our model, equation (\ref{bosons}). The mean field solution produces very interesting patterns
for the transition lines, as shown in Fig. \ref{fig:lobes MF}. For details of this calculation, see Methods. \\
We note that the phase diagrams for conditional-hopping bosons and for anyons are the same. Due to the unitarity of the mapping (\ref{mapping}), the two models (\ref{AHM}) and (\ref{bosons}) are isospectral, and thus feature the same energy gaps and phase diagrams.

\begin{figure} []
  \centering
  \includegraphics[width=1.08\linewidth]{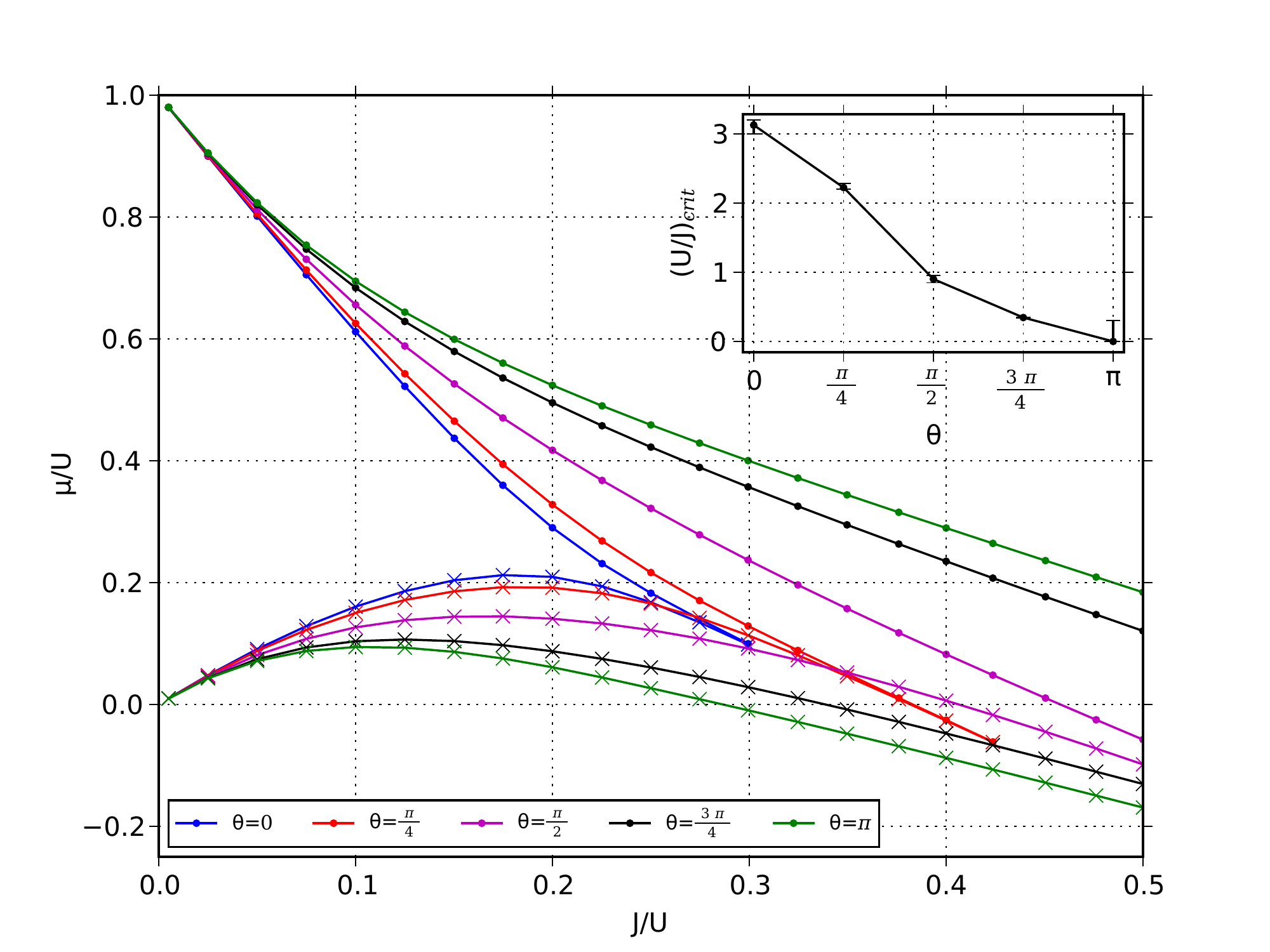}
  \vspace{-0.5cm}
  \caption{
  \textbf{Phase diagram of the anyonized gas.} The Mott lobe (corresponding to $\braket{n}=1$) expands with increasing statistical angle $\theta$ in both directions in the $(\mu, J/U)$-plane. This demonstrates the novel possibility to induce a quantum phase transition from the superfluid into the insulating, Mott-like phase, by simply changing the particle statistics. The energies $\Delta E^{\pm}$ are symbolized by circles (crosses), respectively. The critical phase transition points $(U/J)_{crit}$ are shown on the inset, as a function of $\theta$. In the extreme limit $\theta \rightarrow \pi$, $(U/J)_{crit}$ tends to zero, i.e. pseudofermions seem to be always in the insulating phase.
}
  \label{fig:phases}
\end{figure}

\subsection*{Discussion}

We envision the following procedure to demonstrate the first \emph{statistically induced phase transition}:
We fix the parameters at $J/U = 0.5$, $N/L = 1$ and $\mu \simeq 0$. We start with a phase detuning $\theta=0$ between the external driving fields (see Methods), and thus realize a superfluid bosonic gas as the ground state. The phase angle $\theta$ is now continuously increased, leading to anyonization of the gas and growth of the Mott-like phase. At a critical value $\theta_c$ the phase border will surpass the fixed point in parameter space, which will be then located inside the Mott phase. The critical angle can be estimated from the phase diagram to be in the range $\theta_c \in [\pi/2,3\pi/4]$. For laser detunings beyond this critical angle, $\theta > \theta_c$, the gas will be in an insulating, Mott-like state. In this way, the variation of the particle statistics in our proposal directly realizes a novel superfluid-to-Mott quantum phase transition.

An intriguing aspect of the Mott-like state emerges when a harmonic trapping potential is added to the system, $H^b_{tr} = H^b + V \sum_i ((L+1)/2-i)^2 n_i$. This simulates the experimental conditions also to a more realistic extent. In a harmonic trap, ordinary Mott insulators form real-space density distributions $\braket{n_i}$ that resemble ``wedding cakes'' \cite{BHM}. Due to the broken translational invariance (induced by the trap), the chemical potential in the local density approximation now is a function of the lattice sites. The distribution $\braket{n_i}$ thus exhibits plateaus at integer densities, for ordinary bosons in a Mott-insulating state.

\begin{figure} []
  \centering
  \includegraphics[width=1.02\linewidth]{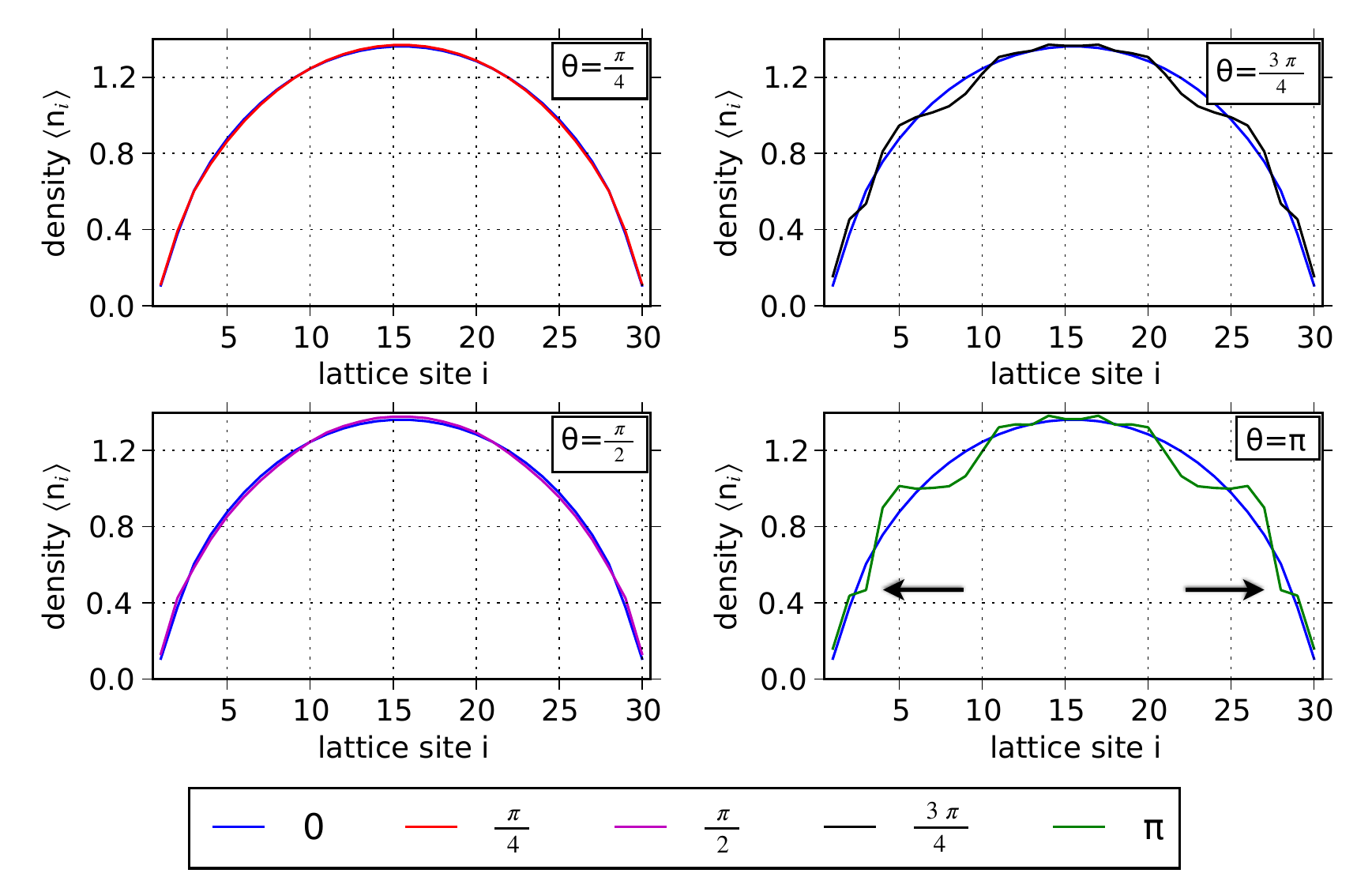}
  \vspace{-0.5cm}
  \caption{
  \textbf{Fractional Mott plateaus.} The density distributions $\braket{n_i}$ are plotted for increasing statistical angle $\theta$.  In each panel the density distribution for ordinary bosons ($\theta=0$, blue line) is drawn for reference. The distribution at high statistical angles ($\theta=\frac{3\pi}{4}, \pi$, marked black and green) exhibit a Mott-like plateau at the fractional value $n \simeq \frac{1}{2}$ (marked by arrows). Parameters: $N=L=30$, $J/U=0.5$, $V/U=0.01$.
}
  \label{fig:mott}
\end{figure}

We have computed the density distribution for a system with additional harmonic confinement, at the same fixed parameter point as in the procedure outlined above. The result is plotted in Fig. 4. For low statistical phase angles $\theta$ (and thus a superfluid state), the density distribution displays a smooth, quadratic profile centered around the minimum of the trap.
Beyond the critical phase angle, $\theta > \theta_c$, Mott-like plateaus appear in $\braket{n_i}$. Surprisingly, in addition to the integer density Mott plateau at $n=1$, a new plateau emerges at the fractional value $n \simeq \frac{1}{2}$. Fractional Mott plateaus persist also in other parameter regimes, when $J/U$ and the trapping potential $V$ are varied, and seem to form a universal feature of 1D anyonic gases at large exchange phases $\theta$, which will be subject of a future study.\\
Recent progress on the experimental side have made direct imaging of density distributions possible, using ``quantum gas microscopy''\cite{Microscope1, Microscope2}. A few weeks ago, the plateau structure of the Mott insulator was directly observed for the first time\cite{Motti} at the single atom level. 
This new technique opens up the possibility to directly demonstrate fractional Mott plateaus and to image statistically induced phase transitions \emph{in situ}.

In summary, we have shown how to realize fractional statistics in 1D optical lattices, using bosons in a realistic experimental environment.
The experiment we propose features the full control and tuneability of the particles' exchange statistics -- paving the way to the first \emph{statistically induced} quantum phase transition.

\subsection*{Methods}

\subsubsection*{Realizing conditional-hopping bosons}

In this section we discuss how to realize the conditional-hopping Hamiltonian equation (\ref{bosons}), using four different and independent $\Lambda$-transitions\cite{Scully}.
In general, we assume a deep optical lattice potential, giving rise to a negligible bare kinetic tunneling amplitude  $J_{kin}$. 
Let us focus for a moment on one $\Lambda$-scheme, where
two ground state levels $|a\rangle$ and $|b\rangle$ are coupled through
a Raman process via an excited state $|e\rangle$.
In our case, $|a\rangle$ and $|b\rangle$ correspond to the wavefunctions
of atoms at distance $d$ localized in neighbouring sites of
the tilted lattice $V(x)$ ($d$ being the lattice constant), while $|e\rangle$ experiences another potential
$V'(x)$ (for a brief discussion of two realistic experimental possibilites to realize $V'(x)$, see the main text).
The levels have energies $E_{i}=\hbar\omega_{i}$,
$i=a,b,e$ and the transition between $a (b)$ and $e$ is driven
by an external radiation field with frequency $\omega_{e}-\omega_{a(b)}-\delta$,
with detuning $\delta$. Note that in our scheme the energies $E_{i}$ depend on $\Delta$ and $U$. The Hamiltonian for the three-level $\Lambda$-system
reads \begin{equation}
H=\sum_{i=a,b,e}\hbar\omega_{i}|i\rangle\langle i|+\frac{\hbar}{2}\left(\gamma_{a}|e\rangle\langle a|+\gamma_{b}|e\rangle\langle b|+\mathrm{h.c.}\right)\label{eq:Ham lasers}\end{equation}
where $\gamma_{a(b)}=\Omega_{a(b)}^{e}W_{a(b)}^{e}e^{-i(\omega_{e}-\omega_{a(b)}-\delta)t}$.
Here, the quantity $\Omega_{a(b)}^{e}$ is the Rabi frequency for the transition
$a(b)\to e$ with the atom centered in the same position. However,
since ground and excited states feel different lattices, the $x$
components of the Wannier functions $w(x)$ are slightly displaced.
Thus, the off-diagonal elements in equation (\ref{eq:Ham lasers}) contain
the integrals\cite{jaksch03} \begin{eqnarray}
W_{a}^{e} & = & e^{ik_{a}x_a}\int w_{e}^{*}(x+x_e)e^{ik_{a}x}w_{a}(x) dx \,,\nonumber \\
W_{b}^{e} & = & e^{ik_{b} (x_a +d)}\int w_{e}^{*}(x+x_e)e^{ik_{b}x}w_{b}(x+d)dx \,,\label{eq:Ws}\end{eqnarray}
where $k_{a(b)}$ is the $x$-component of the momentum of the driving
radiation field, with modulus $|\boldsymbol{\mathrm{k}}_{a(b)}|=(\omega_{e}-\omega_{a(b)}-\delta)/c$. $x_a$ is the atom position in the left well, while $x_e$ refers to the position in the excited state.
Note that the integrals in (\ref{eq:Ws}) are in general non-zero,
as the two Wannier functions belong to different lattices.
The quantities $\gamma_{a(b)}$ are complex numbers, whose modulus
and phase can be freely tuned by choosing the appropriate intensity, polarization
and direction of the driving fields. For sufficiently large detunings $\delta > |\gamma_{a(b)}|$, the
level $|e\rangle$ can be adiabatically eliminated and in the
rotating wave approximation the effective Hamiltonian in the subspace
$\{|a\rangle,|b\rangle\}$ reads
\begin{equation}
H_{\textrm{eff}}=-\frac{\hbar}{4\delta}\left(\begin{array}{cc}
|\tilde{\gamma}_{a}|^{2} & \tilde{\gamma}_{a}^{*}\tilde{\gamma}_{b}\\
\tilde{\gamma}_{b}^{*}\tilde{\gamma}_{a} & |\tilde{\gamma}_{b}|^{2}\end{array}\right) \,,
\end{equation}
where $\tilde{\gamma}$ is the non-rotating version of $\gamma$, i.e. without the time-dependent phase factors.
In order to realize equation (\ref{bosons}), we suggest to employ four driving fields (see labels in Fig. 1b) with different frequencies in order to avoid interferences. This situation would correspond to a maximum of two atoms per site.
If the local density truncation were set to a higher number, more external driving fields would become necessary. As all fields can be tuned independently from each other, this poses no problems besides potential budgetary considerations.\\
The couplings $J_{13}$, $J_{14}$, $J_{23}$ and $J_{24}$, between
the four different levels are then obtained in terms of the effective Rabi
frequencies, $J_{ab}=\tilde{\gamma}_{a}^{*}\tilde{\gamma}_{b}/2\delta$. \\
Our aim is to satisfy the conditions set in equations (\ref{condition1}-\ref{condition2}) in order to engineer a state-dependent phase factor. This implies $\tilde{\gamma}_{2} \equiv \tilde{\gamma}_{1}e^{i\theta}$, which can be achieved by the free tuneability of each driving fields's frequency, intensity, polarization
and direction. Furthermore this choice of parameters implies $|\tilde{\gamma}_{a}| = |\tilde{\gamma}_{b}|$, i.e. the diagonal elements of the effective Hamiltonian are now equivalent. Thus, the tilt energy $\Delta$ has vanished via adiabatic elimination, and consequently also does not appear in equation (\ref{bosons}). Indeed this offset energy between neighbouring sites is compensated by the external radiation field. As the assisted tunneling proposed in this Article is the only mechanism for hopping in the lattice, unwanted effects like Bloch oscillations do not appear in our system (the bare kinetic tunneling amplitude $J_{kin}$ is assumed to be negligible compared to all energy scales discussed here).\\
In summary, the parameters discussed here have to satisfy the following conditions:\\
First, $\delta_{\textrm{linewidth}} \ll \Delta, U$, so that the external driving fields resolve the different levels of the ground state manifold. Second, large detunings $\delta > |\gamma_{a(b)}|$ are required for a short-lived excited state and the validity of the adiabatic elimination. Third, $\Delta$ and $U$ can be in the same frequency regime (a few kHz), but should not be identical (their difference should be $\gg \delta_{\textrm{linewidth}}$). As an example, $\Delta \simeq 2$ kHz, $U \simeq$ 3 kHz, $|J_{ab}| = J \simeq$ 5 kHz  and $|\gamma_{a(b)}| \simeq 20$ kHz would be sufficient if the linewidth of the radiation field were $\delta_{\textrm{linewidth}} \simeq 50$ Hz, which is a realistic assumption for typical radio-frequency driving fields (see e.g. the works by Campbell \emph{et al.}\cite{Campbell} and McKay and DeMarco\cite{McKay}).

\subsubsection*{Fractional Jordan-Wigner mapping}
In the following we prove that anyons are isomorphic to bosons in 1D. \\
In particular we prove that the operators $a$, as defined in the non-local mapping, equation (\ref{mapping}), indeed obey the anyonic commutation relations of equation (\ref{CCR}), provided that the particles of type $b$ are bosons.

For the case $i<j$ we wish to rewrite products of anyonic operators in terms of the bosonic ones:
\begin{eqnarray}
a_i a_j^\dagger &=& b_i e^{-i \theta \sum_{i \leq k < j}n_k} b^\dagger_j \nonumber \\
&=&  e^{-i \theta \sum_{i < k < j}n_k} b_i  b^\dagger_j e^{-i \theta n_i}   \nonumber \, , \\
f(\theta) a_j^\dagger a_i  &=& e^{-i \theta \sum_{i < k < j}n_k} e^{-i \theta n_i} f(\theta) b^\dagger_j b_i \nonumber \\
&=& e^{-i \theta \sum_{i < k < j}n_k} e^{-i \theta (n_i+1)} b^\dagger_j b_i  \, .
\end{eqnarray}
Here we have defined $f(\theta) \equiv e^{i \theta \mathrm{sgn}(i-j)}$ and used that $f(\theta) = e^{-i \theta}$ since $i<j$ was assumed.
We can now evaluate the LHS of equation (\ref{CCR}):
\begin{eqnarray}
a_i a_j^\dagger - f(\theta) a_j^\dagger a_i &=& e^{-i \theta \sum_{i < k < j}n_k} (b_i  b^\dagger_j e^{-i \theta n_i} - e^{-i \theta (n_i+1)} b^\dagger_j b_i)  \nonumber \\
 &=& e^{-i \theta \sum_{i < k < j}n_k} e^{-i \theta (n_i+1)} [b_i,b^\dagger_j] \nonumber \\
 &=& 0
\end{eqnarray}
Thus the anyonic commutation relations have been proven for the case $i<j$. The proof for the case $i>j$ is very similar.
For the case $i=j$ one just has to note that $a_i^\dagger a_i = b_i^\dagger b_i$ and $f(\theta)=1$.
Note that the resulting conditional-hopping bosonic Hamiltonian, equation (\ref{bosons}), resembles the exactly solvable model of q-bosons\cite{qbosons}. However, this model is not equivalent to our model.

\begin{figure} []
  \centering
  \includegraphics[width=1.0\linewidth]{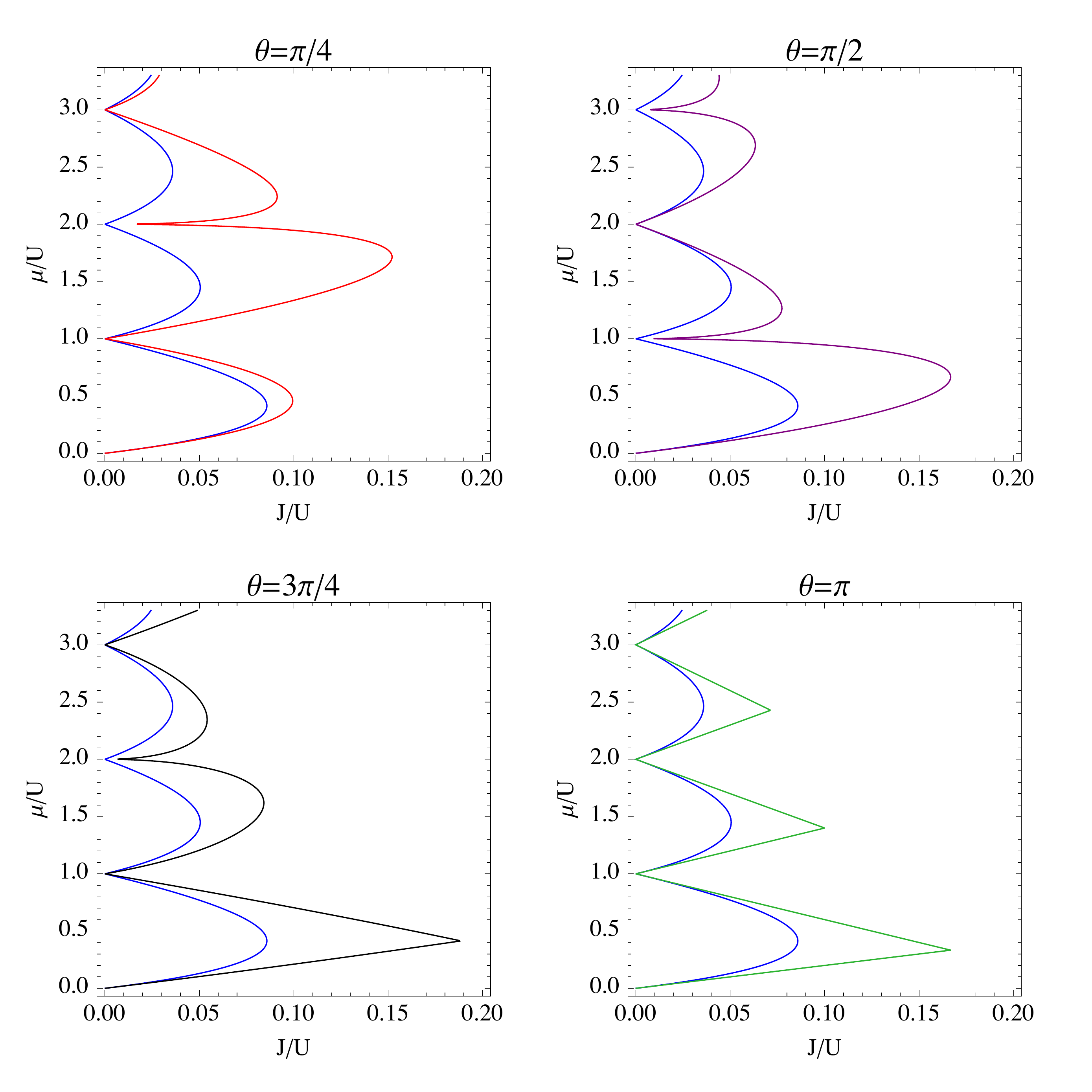}
  \vspace{-0.5cm}
  \caption{
  \textbf{Mean-field solution for the Mott-superfluid transition.}  The transition lines are displayed for several values of $\theta$. For comparison
we display also the data for the bosonic case $\theta=0$ (blue curve).
}
  \label{fig:lobes MF}
\end{figure}

\subsubsection*{Mean-field calculation}

The conditional-hopping bosonic Hamiltonian is given by equation (\ref{bosons}). In addition, we now include also a chemical potential term.
Expressing energies in units of $U$ (we fix $U\equiv1$), we have 

\begin{eqnarray}
H & = & \sum_{j}\left[\frac{1}{2}n_{j}(n_{j}-1)-\mu n_{j}-J(c_{j}^{\dagger}b_{j+1}+b_{j+1}^{\dagger}c_{j})\right]\label{eq:anyon Hubbard}\end{eqnarray}
where for convenience we have defined $c_{j}=e^{-i\theta n_{j}}b_{j}$. 

In absence of hopping $J=0$, all the sites are independent and the
ground state is of the Gutzwiller type\[
|\Psi_{0}\rangle=|\psi\rangle^{\otimes L},\qquad|\psi\rangle=\sum_{\nu=0}^{\infty}c_{\nu}\frac{(b^{\dagger})^{\nu}}{\sqrt{\nu!}}|0\rangle,\]
where $\nu=N/L$ is the filling factor. The local energies are $\epsilon(\nu)=\frac{1}{2}\nu(\nu-1)-\mu\nu$,
while the local gaps for adding or subtracting one boson are, respectively
\[
\epsilon(\nu+1)-\epsilon(\nu)=\nu-\mu,\qquad\epsilon(\nu-1)-\epsilon(\nu)=-(\nu-1)+\mu.\]
So, the ground state in every site has $\nu$ particles in the interval $\mu_{-}^{(\nu)}<\mu<\mu_{+}^{(\nu)}$
, with $\mu_{-}^{(\nu)}=\nu-1$ and $\mu_{+}^{(\nu)}=\nu$. The gap
in the whole system at a given number of particles is given by $\Delta=1$,
obtained by removing an atom at some site an putting it in another
one already occupied. 

Here the mean field (MF) is obtained by decoupling the hopping term
as $c_{j}^{\dagger}b_{j+1}\approx-\alpha_{2}^{*}\alpha_{1}+\alpha_{2}^{*}b_{j+1}+\alpha_{1}c_{j}^{\dagger}$,
where the order parameters are $\alpha_{1}=\langle b_{j}\rangle$
and $\alpha_{2}=\langle c_{j}\rangle$. Accordingly, the Hamiltonian
(\ref{eq:anyon Hubbard}) in MF becomes 

\[
H=\sum_{j}H_{j}+LJ(\alpha_{2}^{*}\alpha_{1}+\alpha_{1}^{*}\alpha_{2})\]
with\[
H_{j}=\frac{1}{2}n_{j}(n_{j}-1)-\mu n_{j}-J\sum_{j}(\alpha_{2}b_{j}^{\dagger}+\alpha_{2}^{*}b_{j}+\alpha_{1}c_{j}^{\dagger}+\alpha_{1}^{*}c_{j}).\]
The two parameters $\alpha_{1}$ and $\alpha_{2}$ are not completely
independent as as they are both vanishing or non vanishing at the
same time. Of course there is a trivial solution corresponding to
$\alpha_{l}=0$, $l=1,2$, which corresponds to the Mott insulating phase. The
occurrence of $\alpha_{l}\neq0$ signals the instability towards superfluid
correlations. On inhomogeneous lattices a further situation can in
principle occur, where $\alpha_{l}\neq0$ only on a fraction of the
lattice sites. The self-consistent relation defines a map $\alpha_{l}=\Lambda_{ll'}\alpha_{l'}$,
obtained by linearizing about the solution $\alpha_{l}=0$. The instability
of the trivial solution sets in when the maximal eigenvalue of $\Lambda$
is greater than 1. Close to the trivial point, it holds $|\alpha_{l}|\ll1$,
hence the kinetic term can be treated perturbatively. Up to first
perturbative order, the wavefunction can be written as $|\psi\rangle=|\psi^{(0)}\rangle+|\psi^{(1)}\rangle$,
where $|\psi^{(0)}\rangle=|\nu\rangle$ and \begin{eqnarray*}
|\psi^{(1)}\rangle & = & -J\sum_{\nu'}\frac{\langle\nu'|\alpha_{2}b_{j}^{\dagger}+\alpha_{2}^{*}b_{j}+\alpha_{1}c_{j}^{\dagger}+\alpha_{1}^{*}c_{j}|\nu\rangle}{\epsilon(\nu)-\epsilon(\nu^{\prime})}|\nu^{\prime}\rangle\\
 & = & J\frac{\sqrt{\nu}(\alpha_{2}^{*}+\alpha_{1}^{*}e^{-i\theta(\nu-1)})}{\mu-\nu+1}|\nu-1\rangle\\
 &  & +J\frac{\sqrt{\nu+1}(\alpha_{2}+\alpha_{1}e^{i\theta\nu})}{\nu-\mu}|\nu+1\rangle\end{eqnarray*}

Hence using the self-consistency relations $\alpha_{1}=\langle\psi|b_{j}|\psi\rangle$
and $\alpha_{2}=\langle\psi|c_{j}|\psi\rangle$, \begin{eqnarray*}
\alpha_{1}/J & = & \frac{\nu(\alpha_{2}+\alpha_{1}e^{i\theta(\nu-1)})}{\mu-\nu+1}+\frac{(\nu+1)(\alpha_{2}+\alpha_{1}e^{i\theta\nu})}{\nu-\mu}\\
\alpha_{2}/J & = & \frac{\nu(\alpha_{2}e^{-i\theta(\nu-1)}+\alpha_{1})}{\mu-\nu+1}+\frac{(\nu+1)(\alpha_{2}e^{-i\theta\nu}+\alpha_{1})}{\nu-\mu}\end{eqnarray*}
The matrix $\Lambda$ is then \[
\Lambda=J\left(\begin{array}{cc}
f(\theta) & A\\
A & f(-\theta)\end{array}\right),\qquad f(\theta)=e^{i\theta\nu}\left[A+(e^{-i\theta}-1)B\right]\]
with \[
A=\frac{\mu+1}{(\mu-[\mu])([\mu]-\mu+1)},\qquad B=\frac{[\mu]+1}{\mu-[\mu]},\]
since every lobe is labelled by $\nu=[\mu]+1$. The eigenvalues of
$\Lambda$ are given by 

\[
\lambda_{\pm}=\frac{J}{2}\left[f(\theta)+f(-\theta)\pm\sqrt{4A^{2}+(f(\theta)-f(-\theta))^{2}}\right].\]
The condition $\max\{\lambda_{+},\lambda_{-}\}=1$ signals the onset
of instability of the trivial solution and hence establishes the critical coupling
$J_{crit}$ along the Mott-superfluid transition line. In Fig. \ref{fig:lobes MF} the phase diagrams for several values of $\theta$ are shown. The Mott lobes expand for non-zero statistical angles $\theta$, a fact central for designing statistically induced phase transitions.

\textbf{Acknowledgements}
We thank Ignacio Cirac, Stefan Trotzky, Ulrich Schollw\"ock and Maciej Lewenstein for helpful discussions. T.K. acknowledges funding from both LMU and Wellness Heaven.

\textbf{Contributions}
The theory and the experimental proposal were conceived by T.K. and M.R.  Numerical data analysis was performed by S.L. under supervision of T.K. The DMRG algorithm was provided by I.M. The manuscript was written by T.K. and M.R.

\textbf{Competing financial interests}
The authors declare no competing financial interests.

\textbf{Corresponding author}
Correspondence to: T. Keilmann (t@ssilo.net)

\end{document}